\documentclass{elsart}
\usepackage{graphicx}
\usepackage{amssymb}

\begin{document}

\begin{frontmatter}

\title{Protein crystallization {\it in vivo}}
\author{Jonathan P.~K.~Doye}
\address{Department of Chemistry, University of Cambridge, 
Lensfield Road, Cambridge CB2 1EW, United Kingdom} 
\author{Wilson C. K. Poon}
\address{SUPA and School of Physics, The University of Edinburgh, Edinburgh EH9 3JZ, United Kingdom}

\begin{abstract}
Protein crystallization in vivo provides some fascinating examples of biological 
self-assembly. Here, we provide a selective survey to show the diversity of functions
for which protein crystals are used, and the physical properties of the crystals that
are exploited. Where known, we emphasize how the nature of the protein-protein interactions
leads to control of the crystallization behaviour.
\end{abstract}

\end{frontmatter}

\section{Introduction}
\label{sect:intro}
The crystallization of proteins {\it in vitro} is a subject of immense practical importance, partly because of the vital role played by protein crystallography in modern structural molecular biology. 
Thus, considerable effort has been devoted to understanding how to crystallize proteins in the laboratory \cite{McPherson99}. 
In the last decade, colloid scientists have contributed significantly to this enterprise \cite{Vekilov02}. 
By treating globular proteins in a coarse-grained manner, it turns out that certain regularities, such as the existence of an optimal `crystallization window', can be rationalised \cite{Poon97}. 
Initially, the simplest possible model was used --- complex protein molecules were reduced to perfect hard spheres with isotropic, short-range inter-particle attraction. Subsequently, more realistic models begin to appear. 
For example, the `stickiness' can occur in patches (either regular \cite{Sear99c} or random \cite{Asherie02}). 
On the other hand, the study of proteins can give new insights into problems in colloid science, such as the existence of finite, equilibrium clusters in charged, attractive particle systems \cite{Poon04}. 
Thus, the intellectual traffic between the study of protein solutions and colloidal suspensions has proved fruitful for both sides.

Protein crystals also occur {\it in vivo}. This phenomenon is, of course, one example of biological self-assembly. Some other examples include the formation of viral capsids \cite{Casjens75} and multi-protein complexes such as the ribosome \cite{FromontRacine03}. Compared to these more celebrated examples, {\it in vivo} protein crystallization has been relatively neglected: we know of only one attempt to provide a general survey (see the last chapter in \cite{McPherson99}). One of the reasons for this relative neglect is that {\it in vivo} protein crystallization is perceived to be an atypical behaviour, and for good reason: protein aggregation and crystallization would normally be expected to be deleterious to the cell. Indeed, the difficulty of crystallizing proteins {\it in vitro} may reflect precisely the `negative selection' that has occurred {\it in vivo} over evolutionary time scales against easily crystallizable variants \cite{Doye04b}. 

The purpose of this article is to provide a review of {\it in vivo} protein crystallization from the perspective of colloid science. Unlike other articles in this journal, many of the references will necessarily not be `current' in the chronological sense --- some of the first examples of {\it in vivo} protein crystals go back nearly a century or more. However, we hope that the topic is of significant current {\it interest}. Understanding {\it in vivo} protein crystallization, and perhaps how cells have evolved to avoid it in the main, may be an area that is ripe for contributions from colloid scientists. We organise the examples reviewed according to the putative biological functions of the protein crystals {\it in vivo}. In the concluding discussion we seek to draw together some elements of commonality between different kinds of {\it in vivo} protein crystals, between {\it in vivo} and {\it in vitro} protein crystallization, as well as point to some of the interesting differences. 

\section{Protein storage}

Perhaps the most `obvious' use for protein crystals {\it in vivo} is for storage, either temporarily, with a view to future utilization or excretion, or as a mean of permanent sequestration. This section gathers examples of this kind.

\subsection{Seeds}

It is instructive to start our review with seeds, since the occurrence of {\it in vivo} protein crystals here is, at least at first sight, rather `obvious' from a physical sciences point of view. A major function of a seed is to act as a storage organ, of carbohydrates, and, in some cases, of proteins. Seed proteins were amongst the first to be crystallized in the laboratory \cite{McPherson99}. Given the ease of {\it in vitro} crystallization, and the fact that many seed proteins are present at very high concentrations, one may expect the presence of protein crystals {\it in vivo}. This expectation is amply confirmed by reality: the packing within membrane-bound storage organelles, `protein bodies' (previously) or protein storage vesicles (in more recent literature), often contain crystalline portions. Indeed, protein crystals in seeds were first reported as early as 1855, and so is one of the first known examples of {\it in vivo} crystallization. Perhaps the most striking demonstration of crystallinity comes from diffraction from actual seeds. In particular, dry and wet slices of seeds from rock melon and pumpkin were found to shown powder diffraction rings consistent with a surprisingly simple lattice (face-centred cubic) \cite{cucurbitin}. 

In more detail, the protein storage vesicles in most non-legume seeds consist of three parts: an amorphous matrix, `crystalloids' embedded in the matrix containing proteins packed in lattice structure, and crystals of the salts of phytic or oxalic acids (see, e.g. \cite{Lott80}, especially the schematic drawing in Figure 1 of this review, and Figures 10 and 11 showing lattice steps in crystalloids). The detailed mechanism whereby such complex storage organelles are constructed is still an active area of research \cite{Muntz,Jiang00}. It is clear that understanding why some of the proteins are sequestered into crystalline compartments (e.g., is it some form of micro-phase separation?) will be a crucial part of this enterprise. 

In common with other legumes, soya beans show no evidence of crystalloids. Nevertheless, one of the soya storage proteins provides a surprising example of {\it in vivo} crystallinity when expressed in transgenic wheat \cite{Casey}. Since the domestication of {\it Glycine max} in northern China more than 3000 years ago, soya beans have been a major source of dietary protein for a substantial fraction of the world's population. The two main soya storage proteins are glycinin and $\beta$-conglycinin (often also referred to by their sedimentation coefficients as 11S and 7S) \cite{Boulter}. Like many other seed proteins, the wild type of both soya proteins are easy to crystallise. Nevertheless, their atomic-resolution structures have remained illusive until recently \cite{conglycinin,glycinin}, because crystals from the wild-type protein do not give diffraction patterns of sufficient quality due to `multimer heterogeneity'. Here we concentrate on glycinin.

Glycinin is a hexamer, each monomer of which is composed of an acidic and a basic subunit. Each pair is synthesized as a single precursor (but later cleaved). Three of these monomers are first assembled into a trimer in the endoplasmic reticulum. The trimer is transported to the storage vesicle, further proteolytically modified, and then assembled into the final hexamer. The monomer subunits are produced from multiple genes, with between 50-90\% sequence identity between them. Mixtures of these monomers assemble to give rise to heterogeneous hexamers, which cannot give high-resolution diffraction patterns. One strategy to overcome this problem is to produce recombinant glycinin in transgenic wheat containing a single type of subunit. In such transgenic wheat, the glycinin is observed to adopt a lattice structure when the seed develops \cite{Casey}. Thus, it appears that modifications that favour crystalline packing (here, molecular homogeneity) but do not otherwise hinder function do indeed lead to crystallization {\it in vivo}. 

\subsection{Secretory granules}

Many cell types are involved in regulated secretion, in which particular
proteins are synthesised in significant quantities, and then stored in concentrated
form ready for release from the cell in response to some external signal.
On synthesis proteins destined for secretion enter into the rough
endoplasmic reticulum (RER), and then pass through the Golgi apparatus, where 
proteins are sorted according to their destination. Vesicles with the
relevant protein in a condensed state form at the trans-Golgi network, and then further  densify to form the mature secretory granules \cite{Arvan98}. 
The proteins in these granules are probably more often in a dense amorphous state, but there are a significant number of examples where they are crystalline. When secretion is  triggered, the vesicles migrate to the cell surface, and release their
contents by exocytosis. The protein granules then dissolve.

In order for this secretory pathway to be successful,
the cell must exert considerable control over the protein's interactions
through control of the environment experienced by the protein. 
Otherwise premature aggregation or crystallization, say in the RER \cite{Arias93}, 
might occur or the granule might fail to dissolve on exocytosis.
As one passes through the RER and Golgi apparatus, the pH decreases and 
the concentration of divalent ions, such as Ca$^{2+}$ and Zn$^{2+}$ 
changes significantly.
Indeed, aggregation has been found to occur {\em in vitro} for 
a series of secretory proteins under conditions that mirror those
of the trans-Golgi cisternae \cite{Dannies99}.
Also, it is fairly common for the secretory protein to be initially
synthesised as a proprotein, which is then proteolytically processed in the 
Golgi apparatus to produce the active form.
This provides another avenue to control
the protein's aggregation and crystallization properties.

It has also been suggested that the aggregation might play an important 
role in the sorting of secretory proteins in the Golgi apparatus by a mechanism 
called ``selection by retention'' \cite{Arvan98,Dannies99}. 
Lysosomal proteins and those that are part of the constitutive secretory pathway 
have a sorting signal in their amino acid sequence and are actively trafficked
away from the Golgi apparatus to their correct destination. 
The essence of the proposed mechanism is that what remains 
ends up in the regulated secretory pathway. 
It has been suggested that, as well as the absence of a sorting 
signal, the formation of insoluble aggregates prevents these proteins from entering
into vesicles heading to these alternative destinations.

\subsubsection{Insulin}
One of the most well-known examples of regulated secretion is that of insulin,
and one for which there is a good understanding of the structural changes
associated with the secretory process \cite{Dodson98}.
Granules of insulin are secreted from the islets of Langerhans in the pancreas into 
the blood in response to elevated glucose levels in the blood. 
The granules are typically 200--300$\,$nm in size, and contain insulin in a crystal form.
Insulin and proinsulin exists in a number of oligomeric states, and
this is exploited in the secretory process. The initially synthesised proinsulin
assembles into a soluble hexamer in the presence of zinc irons. Conversion 
of proinsulin to insulin by cleavage of the `C peptide' from the A and B chains occurs
during formation of the secretory vesicle. The insulin hexamers are insoluble in the presence of zinc ions and crystallization occurs rapidly. 

The function of the crystalline form
is not clear, but it may serve to protect the insulin from further proteolytic
processing. 
Interestingly, in species such as the guinea pig, where insulin is not hexameric, 
dense amorphous aggregates occur instead \cite{Dodson98}. 
In this context, it would be interesting to compare the rate of proteolysis on 
the differently aggregated forms of insulin {\it in vitro}. 

On exocytosis of the insulin granules, there
is a significant increase in pH and a decrease in concentration of Zn$^{2+}$ and Ca$^{2+}$.
These changes lead to the deprotonation of six glutamate residues in close proximity at the 
centre of the hexamer, and the loss of the coordinating zinc ions. The resulting electrostatic
repulsion causes the crystals to dissolve rapidly to give the functionally active monomeric
form of insulin.
Given this context, it is unsurprising that insulin is one of the most rapidly 
crystallizing proteins {\em in vitro}.

\subsubsection{White blood cells}
Granulocytes are a type of white blood cell characterized by secretory granules packed with
potent chemicals that can be released to combat infection. Some of the proteins in
these granules
are stored in the form of crystals. For example, in the eosinophils
depicted in Fig.\ \ref{fig:montage}(a), 
large rectangular crystals consisting of the eosinophil major basic
protein (EMBP) are apparent \cite{Giembycz99}. 
This protein is synthesised as pro-EMBP, where the
pro-portion of the protein is highly acidic, presumably in order to counteract the 
highly basic nature of EMBP and thus protect the cell during transport to the granule,
where the pro-portion is then removed \cite{Swaminathan01}. 

Eosinophils also have smaller
granules containing Charcot-Leyden crystal protein \cite{Leonidas95}. 
Interestingly, after release of this protein by eosinophils during an inflammatory
response, the high concentrations can lead the protein to recrystallize. 
The resulting hexagonal bipyramidal crystals, known as Charcot-Leyden crystals, 
are characteristic of eosinophil-associated allergic inflammation, such as
asthma, and were first observed in 1851 \cite{Charcot53}. 
Whether this recrystallization serves
any functional purpose is not clear, but possibilities are that it provides
a way to deactivate excesses of this destructive protein,
or that it provides the first step in the removal of the protein from the 
tissue, e.g.\ the subsequent uptake of the crystals by macrophages has been 
observed.

\subsubsection{Protists}
There are many examples of crystalline secretory granules amongst the protists 
\cite{Hausmann78},
a diverse group of mainly single-celled eukaryotic organisms. 
Usually the material extruded from the cell is for 
the purpose of predation of or protection against other 
microorganisms. 
Two of the most well-studied examples are from {\em Tetrahymena} 
\cite{Turkewitz04} and {\em Paramecium} \cite{Vayssie00},
and illustrate the complexity of crystalline secretory granules 
that can be formed. 
The secretory granule of {\em Paramecium}, called a trichocyst, 
(Fig.\ \ref{fig:montage}(b)) 
involves the assembly of at least three different families 
of closely related polypeptides, each localized 
to different regions of the trichocyst to form a structure that is
of the order of 3--4$\,\mu$m in size. These polypeptides are
derived by proteolytic cleavage of proproteins during the development 
of the granule, and the absence of any leads to misassembly of the
trichocyst \cite{Vayssie01}. 
Furthermore, this proteolysis also serves as a
means to control the intermolecular interactions during trafficking, 
since the soluble proproteins can easily pass through the endoplasmic reticulum and 
the Golgi apparatus, whereas the final polypeptides are insoluble.
Therefore, proteolysis triggers aggregation, and the retention of 
these polypeptides in the maturing secretory granules. 
Interestingly, in the absence of this proteolysis, the proproteins end up in
the constitutive secretory pathway \cite{Vayssie00}.

Rapid and synchronized exocytosis in response to external
stimulus is triggered by the release of calcium ions
leading to substantial (roughly by a factor of 8) 
expansion of the crystal as it is extruded. This process is irreversible,
implying that the original assembly of the trichocyst results in a metastable 
structure with energy stored in the crystal lattice.
The purpose seems to be defensive. The explosive release
of the trichocysts can push {\em Paramecium} away from a potential predator, giving
it a chance to escape \cite{Knoll91}.
The physical mechanisms causing the expansion is 
unknown, although it may be relevant that the polypeptides 
involved are all acidic.
Similar large expansions of amorphous polyelectrolyte gels
occur in the secretory granules of mast cells, where the negatively-charged
polymer matrix exhibits a condensed state in the presence of divalent
ions, but undergoes a rapid expansion in the presence of monovalent ions, or
an electric field \cite{Nanavati93}.

\subsection{Bacterial parasporal crystals} 
 
On sporulation, a number of species of bacteria,
the most well-studied of which is {\em Bacillus thuringiensis}, 
form protein crystals containing insecticidal toxins (Fig.\ \ref{fig:montage}(c)).
When ingested, these crystals dissolve in the alkaline environment of the gut
of the insect larvae, releasing the protoxins which are then cleaved to produce 
the active form that attacks the gut wall and facilitates entry of germinating spores into 
the insect host. There is a large family of these toxins, and the crystal 
structures of many of these are known \cite{Schnepf98,deMaagd03}.
It is common for these proteins to be cysteine-rich and it is thought that
intermolecular disulphide bonds contributes to their crystallization 
behaviour. In others, there are strong intermolecular salt bridges. 
Interestingly, when expressed in a variety of other organisms (even plants)
crystal formation can still occur \cite{DeCosa01}. 
The reason for packaging the toxins in a crystal is probably 
to provide a concentrated dose of toxin in a form that is insoluble
at neutral pH, thus increasing the longevity  
of the toxins in the environment of the soil \cite{deMaagd03}.

\section{Encapsulation}

Some particularly interesting examples of protein crystallization 
occur in certain genera of insect viruses, namely cytoplasmic and 
nuclear polyhedrosis viruses, granulosis viruses and entomopoxviruses \cite{Smith76}.
All these viruses coopt the infected cells to express large 
quantities of proteins (polyhedrin, granulin and spheroidin, respectively, 
in the above classes) in the late stages of infection.
(Indeed, the baculovirus protein expression systems, one
of the most popular eukaryotic alternatives to {\em E. coli}, make use
of the relevant promoter in cut-down versions of the {\it Autographa californica} 
nuclear polyhedrosis virus, where the gene for the expressed protein replaces that 
for polyhedrin.)
These proteins then crystallize around the the virions to provide
a protective environment for the viruses after death of the insect larvae,
which can remain intact for periods of years in the soil. 
The viruses also induce cell lysis, aiding release of the crystals 
from the dead insect larvae, and even induce larvae to climb upwards 
(hence the name Wipfelkrankheit or tree-top disease) to aid as wide dispersal 
as possible \cite{Goulson97}.
On ingestion of the crystals by a new insect larva, the protein crystals
dissolve in the alkaline environment of the gut, releasing the virus to 
infect the new host. 

\begin{figure}
\begin{center}
\includegraphics[width=12.3cm]{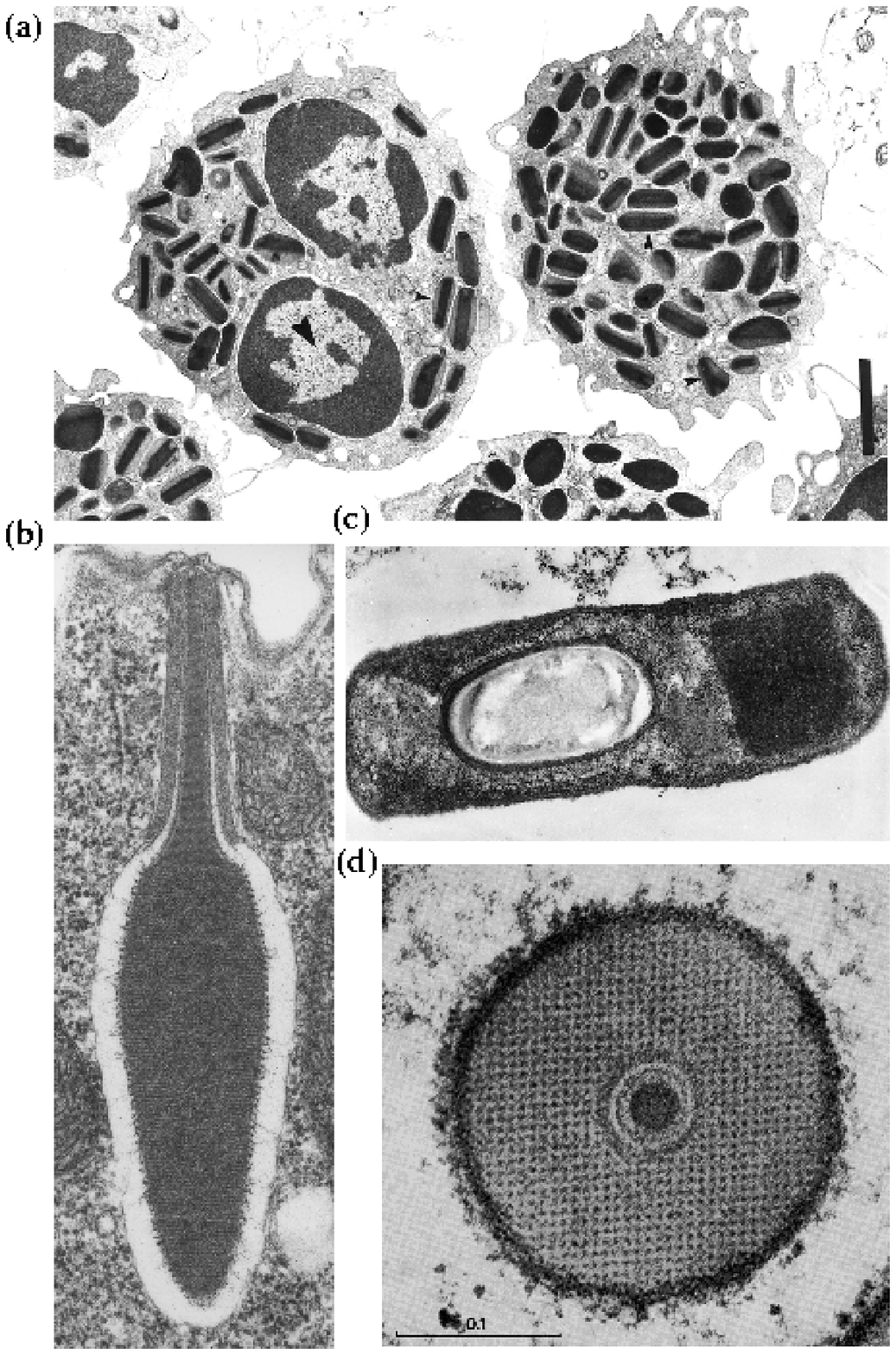}
\end{center}
\caption{
(a) Eosinophils showing granules containing dark 
rectangular crystals of EMBP.
(b) A trichocyst attached to the outer membrane of {\em Paramecium}.
(c) A protein toxin crystal within {\em Bacillus thuringiensis}. 
(d) An encapsulated virus rod of the granulosis virus of {\em Plodia interpunctella}.
Scale bars correspond to (a) 1$\,\mu$m and (d) 0.1$\,\mu$m.
Reproduced with permission from Refs.\ (a) \cite{Giembycz99} (b) \cite{Vayssie00},
(c) \cite{Agaisse95} and (d) \cite{Arnott68}.}
\label{fig:montage}
\end{figure}

Depending on species, the protein crystals can either encapsulate
single or multiple virus particles. The example shown in Fig.\ \ref{fig:montage}(d) 
shows a crystal encapsulating a single rod-like virus particle. 
Particularly, interesting
is the uniformity of the crystals --- the cross-sections of the crystal are almost 
perfectly circular, and the crystals are relatively monodisperse. These
features raise a host of physical questions. How does the crystal nucleate around the
virus, in this case on the surface of the membrane surrounding the virus, particularly as
heterogeneous nucleation usually leads to crystal growth away from the nucleating particle \cite{Cacciuto04}?
What limits the growth of the crystals to a size that is large enough to offer 
sufficient protection to the virus, whilst not using excessive amounts of protein? 
Interestingly, there is some evidence from ultrastructural studies that 
for granulosis viruses, such as in Fig.\ \ref{fig:montage}(d), the nucleation of 
the crystal usually begins at one end of the rod virus, and that the crystal then
grows around the virus \cite{Arnott68}.

The mechanical properties of the crystals are also important for their function. Whereas
most protein crystals are very fragile, in order to protect the virus, these crystal are very
tough, implying very strong interprotein interactions. However, how then are 
the proteins transported to the site of crystallization without any premature aggregation
or crystallization? Indeed, if crystals of polyhedrin produced {\em in vivo}
are dissolved in alkali {\em in vitro} 
(having prevented expression of a protease that co-crystallizes
with polyhedrin, and whose purpose is to speed up dissolution of the crystal 
in the alkaline environment of the gut by cleaving polyhedrin into smaller pieces)
on lowering the pH aggregation rather than crystallization occurs \cite{Chiupersonal}.
Consequently, the crystal structure of these proteins are unknown.
It may be that other highly expressed proteins, such as p10 \cite{vanOers97}, which 
forms filaments that are intimately associated with the polyhedrin crystals 
and is involved in the formation of envelopes around the crystals, 
may also play a role in the transport of polyhedrin. 

\section{Solid-state catalysts: Peroxisome enzymes}

Peroxisomes are membrane-bound organelles found in
eukaryotic cells that are responsible for a variety of chemical processes, 
such as the breakdown of lipids and alcohol, that are catalysed
by enzymes, such as catalase, urate oxidase and alcohol oxidase, that 
are typically assembled into regular crystals. 
These reactions are usually oxidative in nature and often involve 
unpleasant chemical species such as H$_2$O$_2$, hence the need for 
confinement in a specific organelle.
The crystals can be both co-crystals of a mixture of enzymes or a pure
crystal of one enzyme depending on the cell involved, and
typically have an open crystal structure with clearly observable solvent
channels.

The peroxisomes in yeast cells fed on methanol provide a particularly 
well-studied example \cite{Veenhuis03}. 
In this case, the crystals are pure alcohol oxidase, 
and the peroxisomes can take up a majority of the cell volume. Of course,
it is important that the enzymes do not assemble or take on their active 
form prior to entry into the peroxisome. In the case of alcohol oxidase, the
active form is an octamer. After synthesis on free ribosomes in the
cytosol, alcohol oxidase binds to another protein, pyruvate carboxylase (pyc), 
preventing formation of the octamer. Only after entry into the peroxisome and 
release from pyc, which remains in the cytosol, does the protein form
the octamer and spontaneously assemble into a crystal.
Interestingly, in mutants that lack peroxisomes, crystals do not form,
except at particularly high concentrations of alcohol oxidase, when they 
can form in the cytoplasm \cite{vanderKlei04}.

Why are the enzymes organized into crystals? As the molecules that
the enzymes act upon are trafficked to the peroxisome, 
there is no need for the enzymes to diffuse through solution to find 
their reactants. Moreover, as the relevant molecules are not other proteins, 
but are generally relatively small, they can easily diffuse through the open crystals. 
The only requirement is that the active site is not at a crystal contact.
The crystals, therefore, represent nanoporous solid-state catalysts.

\section{Plugging leaks}

\subsection{Woronin bodies} 

In filamentous fungi, cellular compartments
are connected by septal pores that allow trafficking between cells. 
When a cell is damaged, the septal pore is sealed by Woronin bodies, 
preventing cytoplasmic bleeding. Each Woronin body is a hexagonal platelet 
$\approx 5\,\mu$m across, and consists of a proteinaceous core that is primarily 
crystalline HEX-1 surrounded by a membrane \cite{Jedd00}. The HEX-1 protein contains a
peroxisomal targeting sequence that causes it to be trafficked to the organelles,
and thus the Woronin bodies can be classified as peroxisomes, 
although, unlike most peroxisomes, they are not known to carry out a catalytic function. HEX-1 crystallizes readily {\em in vitro} \cite{Jedd00}, 
and its structure has been determined to high resolution \cite{Yuan03}.

Interestingly, a mutation exists in HEX-1 that leads to the formation of aberrant spherical Woronin bodies, taken by the investigators concerned as an indication of non-crystallinity \cite{Yuan03}. These spherical particles fail to seal septal pores about damage to fungal hyphae. It has been suggested \cite{Yuan03} that
the Woronin bodies require a dense core in order to 
withstand the intracellular turgor pressure (which can be as high as 80~MPa \cite{Turgor}) and hence provide an efficient seal. It is not clear, however, that dense amorphous packing would achieve this any less effectively. 

\subsection{Phloem sieve elements} 

In plants, transport in the phloem occurs along sieve 
tubes. The elements of these tubes are connected by sieve plates that contain
large pores. When these cells are damaged, it is important that these pores are 
blocked to prevent leakage. This is achieved through changes 
to various proteinaceous bodies in the elements \cite{vanBel03}. One
of these bodies involves P-proteins in crystalline form. In legumes
these P-protein crystalloids can undergo a reversible transformation
to a dispersed state that can plug the sieve tubes in response to the
presence of Ca$^{2+}$ \cite{Knoblauch01}.

\section{Disease induced crystallization} 

For most proteins, crystallization within the cellular is to be avoided, and 
crystallization is a sign of dysfunction. For example, we know of two diseases that
are directly caused by protein crystallization. 
Firstly, hemoglobin C disease is associated with a mutant form of 
hemoglobin, in which a glutamate is replaced by lysine.
Crystallization of this hemoglobin C can occur in the red blood cells of people 
who are homozygous for the gene encoding this mutation (CC) or
who have a combination of this and the sickle-cell gene (SC). The former case leads 
to hemoglobin C disease, which is a mild form of anaemia. 
The physical mechanisms of crystallization of hemoglobin C 
have been intensively studied \cite{Vekilov02b,Chen04,FeelingTaylor04}.
Secondly, mutations in the $\gamma$-crystallin proteins can lead to cataracts. 
In some of these cases, the increase in lens opacity is due to 
crystallization of the $\gamma$-crystallin \cite{Pande01}.

There are then numerous examples where protein crystals have been discovered
to be associated with diseases in pathology samples \cite{Lee02}, 
but often where the role played
by the crystals---are they the cause, a harmful side-effect or a harmless byproduct
of the disease---and even the protein that has crystallized remain unclear. 
We have already mentioned the Charcot-Leyden crystals associated with the action
of eosinophils. Immunoglobulins, particular the $\kappa$ light chain, are another
set of proteins that have a propensity to form crystals, often in association with 
cancers \cite{Lebeau02}.

\section{Crystallization of larger proteinaceous structures}

There are also many examples of larger proteinaceous bodies forming (para-) crystalline structures {\em in vivo}.
For example, in cells containing large numbers of icosahedral viral particles, crystals are  frequently seen to occur. The resulting iridescence due to Bragg scattering of visible light lies behind the naming of the iridovirus family \cite{Smith76}. It is not clear whether this behaviour has a functional purpose---perhaps the formation of a crystal  maximises the room for more viruses to be produced, or acts like a more primitive means of encapsulation, i.e.\ the viruses in the centre are protected by those on the
outside, or a means for providing a concentrated dose of the virus on transmission.
Crystal-like structures are also seen in rod-shaped viruses, such as tobacco mosaic virus. Other examples include the formation of ribosome crystals in hibernating animals \cite{McPherson99}, and crystalline arrays of cellulose fibres.

Two-dimensional crystals are often frequently seen. This can be of membrane proteins,
such as cytochrome oxidase \cite{Henderson77},
but also of proteins that provide structural rigidity.
For example, crystalline arrays of S-layer proteins provide added stability to the 
cell envelope of many prokaryotes \cite{Sara00}, and the CcmK proteins form hexamers
that are the building blocks for the polyhedral shell of bacterial microcompartments,
such as the carboxysome \cite{Kerfeld05}. 

\section{Conclusion}

From even such a short survey --- we could have provided many more examples --- it 
is clear that protein crystallization {\it in vivo}, while not necessarily 
ubiquitous, is not nearly as esoteric a phenomenon in biology as one might have thought. 
In our selection, we have dwelt on those cases that are the most well-characterized, 
and the function the most clear. However, in a number of instances, particularly 
those associated with storage, although it is clear why a dense state is favoured,
it is less clear why that state needs to be crystalline rather than amorphous.
This is highlighted by the example of insulin granules, which are crystalline
in humans, but amorphous in guinea pigs.
Any selective advantage of the crystalline state is clearly subtle, 
and it might just be that crystallization is {\em incidental}, i.e.\ neutral 
evolution has led to two equally viable ways of storing insulin.
In such cases, investigating crystallization may yield little fruit for biology, 
except that the study of crystal contacts may serve to corroborate information 
obtained from other means regarding the structural consequences of mutations. 

By contrast, in cases such as the Woronin bodies, the encapsulated viruses, and
the trichocysts, where the mechanical properties of these bodies is vital to
their function, understanding crystallization should also yield dividends 
for understanding essential biological function. Although crystal formation 
might not be necessary {\em per se} to achieve these properties, the 
intermolecular interactions that induce crystallization also perform
important roles in their functions.
Biophysical measurements of the mechanical properties of these protein crystals, 
such as the strength of the HEX-1 and polyhedrin crystal or the forces 
generated by the expansion of the trichocysts, may also have biological pay-offs.

Probably the biggest difference, between protein crystallization {\em in vitro}
and {\em in vivo} is that the location of the {\it in vivo} crystals is vital
to their function. In particularly, the cell needs to transport the proteins to
their destination without premature crystallization or aggregation. To achieve
this, the cell exerts an exquisite control over the interprotein interactions,
that would be the envy of any protein crystallographer.
In some of the examples, we have seen how this can be achieved through
changes in the ionic environment, through proteolysis of precursor proteins, and
binding partners that ``chaperone'' the proteins. 

Furthermore, the architecture of the {\em in vivo} crystal, 
both its size and shape, can be vital to its function. This is exemplified
by the complexity of the trichocysts. However, the mechanisms by which
such control over the crystallization and self-assembly is achieved are still 
unknown.  Much can potentially be learnt from these examples by those who 
would like to design and build nanostructures.
Finally, {\em in vivo} crystals are not simply passive. 
Rather, they can assemble and disassemble in response to environmental signals, 
they can resist mechanical stresses, and
even rapidly undergo massive shape changes.

\subsection*{Acknowledgements} We thank Rod Casey and Clare Mills for introducing us to seed storage proteins, Nick Read for educating us about Woronin bodies, 
Ard Louis, Robert Possee and Michele Vendruscolo for stimulating discussions,
and Linda Sperling for provision of original micrographs and for 
helpful comments on the manuscript.

\end{document}